\documentclass[aps,prb,twocolumn,groupedaddress,showpacs]{revtex4}

\usepackage{dcolumn}
\usepackage{bm}
\usepackage[mathlines]{lineno}

\usepackage{amsmath}    
\usepackage{graphicx}   
\usepackage{verbatim}   
\usepackage{color}      
\usepackage{subfigure}  
\usepackage{hyperref}   
\usepackage{algorithm}  
\usepackage{algorithmic}

\begin{document}


\title{Electronic transport of a large scale system studied by renormalized transfer matrix method: application to armchair graphene nanoribbons between quantum wires}
\author{Miao Gao$^{1}$}
\author{Gui-Ping Zhang$^{1}$}\email{bugubird$_$zhang@hotmail.com}
\author{Zhong-Yi Lu$^{1}$}\email{zlu@ruc.edu.cn}
\affiliation{$^{1}$Department of Physics, Renmin University of
China, Beijing 100872, China}

\date{\today}

\begin{abstract}
Study on the electronic transport of a large scale two dimensional system by the transfer matrix method (TMM) based on the Sch\"{o}rdinger equation suffers from the numerical instability. To address this problem, we propose a renormalized transfer matrix method (RTMM) by setting up a set of linear equations from $U$ times of multiplication of traditional transfer matrix ($U=\frac{N}{S}$ with $N$ and $S$ being the atom number of length and the transfer step), and smaller $S$ is required for wider systems. Then we solve the above linear equations by Gauss elimination method and further optimize to reduce the computational complexity from O($U^3M^{3}$) to O($UM^{3}$), in which $M$ is the atom number of the width. Applying RTMM, we study transport properties of large scale pure and long-range correlated disordered armchair graphene nanoribbon (AGR) (carbon atoms up to $10^{6}$ for pure case) between quantum wire contacts. As for pure AGR, the conductance is superlinear with the Fermi energy and the conductance is linear with the width while independent of the length, showing characteristics of ballistic transport. As for disordered AGR with long-range correlation, there is metal-insulator transition induced by the correlation strength of disorder. It is straightforward to extend RTMM to investigate transport in large scale system with irregular structure.



\end{abstract}

\pacs{72.10.-d, 73.21.-b, 73.43.-f, 73.50.Fq}

\maketitle

\section{Introduction}

Transfer matrix method (TMM) based on the Sch\"{o}rdinger equation is a widely used numerical approach to investigate electronic transport, such as in disordered systems \cite{mackinnon,Zhang-1,Zhang-2} or in the presence of electron-phonon interaction \cite{Zhang-3,Zhang-4}. However, when the spacial dimension is higher than 1, the size of a system investigated by TMM is very limited due to numerical instability\cite{mackinnon}. This numerical instability originates from such an issue that the smallest eigen-mode in a considered system will be lost in computation when its ratio to the largest is less than the accuracy of our computer, represented by floating point numbers. Thus it more readily occurs for a wider two-dimensional system that has more eigen-modes and then larger difference between the smallest and largest eigen-modes \cite{reorthnorm,MyReview-01}, especially their ratio dramatically decreasing exponentially with $n$ after $n$ times recursive multiplication of the matrix transfer. To deal with such a numerical instability and realize a large scale calculation, a number of schemes have been proposed, for example, by introducing extra auxiliary parameters that are determined together with reflection coefficients \cite{Yin}, or by diagonalizing the transfer matrix of a conductor using eigenstates of leads\cite{Hu}. So far these schemes only made certain improvement and cannot handle inhomogenous systems yet.

In this article, we propose a renormalized transfer matrix method (RTMM) to calculate the conductance of a large system, meanwhile readily incorporating disorders and/or impurities. We sketch it as follows. Conventionally one recursively multiplies the transfer matrix in a scattering region from one lead side into the other lead side to directly resolve the reflection and transmission coefficients. Here we first divide the scattering region into $U$ subregions, similar to the idea proposed in Ref. \cite{subdivision}. In all the subregions, we respectively take the recursive multiplications of the corresponding transfer matrixes without the numerical instability, and then lump them into a set of linear equations containing the wavefunction values at all the interfaces between the subregions as the unknowns, among which the reflection and transmission coefficients are related to wavefunction values at the left and right lead-scattering interfaces respectively. We then solve this set of linear equations by using a modified Gaussian elimination method, which has been elaborately optimized by us to reduce the computational complexity from O($U^3M^{3}$) ($M$ being the site number of the width) to O($UM^{3}$) loop executions so that a system with a million of lattice sites can be calculated on a standard desktop computer. For a wider system, clearly a larger $U$ is required to avoid the numerical instability. We will illustrate the method by using it to study the electronic transport of graphene in this article.


The discovery of graphene in 2004 \cite{exp0}, a single atomic layer of graphite with carbon atoms sitting at a honeycomb lattice, has aroused widespread interest both theoretically and experimentally, due to its distinctive electronic structure, whose low energy excitations can be interpreted in analogy to massless Dirac relativistic fermion model, and its great potential on practical applications \cite{exp1,exp2,exp3}. Among various graphene-based materials, armchair graphene nanoribbon (AGR) attracts intensive attention since there is an energy gap opened \cite{AGR-gap}. Here we choose AGR as a model system to study.

\begin{figure}[tbh]
\begin{center}
\includegraphics[width=8.5cm]{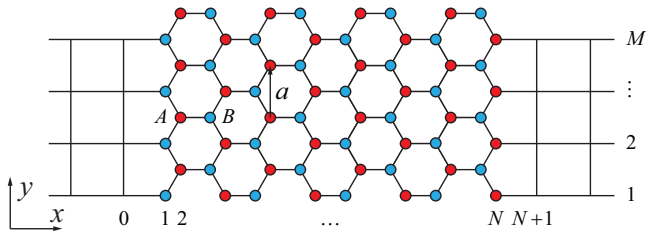}
\caption{(Color online) Schematic of armchair graphene ribbon (AGR) with $N$ and $M$
atoms in $x$ and $y$ direction, as connected to square-lattice leads
simulating normal metal leads. Here the length of the samples are $(3N-4)\sqrt{3}a/12$
and the width $(M-1/2)a$, where $a$ is the lattice parameter of graphene.
Red and blue colors donate different sublattices A and B, respectively.}
\label{fig:structure}
\end{center}
\end{figure}

In most theoretical and numerical studies on the electronic transport of AGR, the leads were made of doped graphene\cite{exp10,exp11}, however, experimentally the leads were usually made of normal metals such as gold\cite{exp0,exp1,exp8,exp9}. Similar to Ref. \onlinecite{Schomerus}, here we employ two semi-infinite square lattice quantum wires as leads to simulate normal metal leads, as shown in Fig.~\ref{fig:structure}. Actually we had previously calculated the transport properties of graphene nanoribbons between such quantum wire leads by the conventional transfer matrix method \cite{ZhangGP-01,ZhangGP-02,ZhangGP-03,disorder-correlation}, which however would be better to be further examined by large scale graphene calculations, especially considering the effect of long-range correlated disorder and/or impurities that has an important impact on the formation of electron-hole puddles observed in graphene\cite{electron-hole-puddles}, as discussed in Ref. \onlinecite{disorder-correlation}. In addition, large scale system calculations are also required to determine whether or not the existence of Anderson localization \cite{Anderson-PR1958} in low dimensional disordered system. Meanwhile, the renormalized transfer matrix method can be readily extended to investigate transport in a large scale graphene system with irregular structure.

In this article we mainly present the renormalized transfer matrix method. We organize the paper as follows. In Section II, we present the renormalized transfer matrix method in conjugation with a tight binding model to describe graphene; in Section III, we introduce optimized Gaussian elimination algorithm; in Section IV, we apply the renormalized transfer matrix scheme to investigate the transport properties of pure armchair graphene nanoribbons and long-range correlated disordered armchair graphene nanoribbons, respectively; and in Section V, we make a summary.

\section{Tight-binding model and renormalized transfer matrix method}

Graphene takes a honeycomb lattice with two sites per unit cell, namely consisting of two sublattices $A$ and $B$. The tight-binding Hamiltonian considering that $\pi$ electrons hop between the nearest-neighbor atoms in graphene reads as follows,
\begin{equation}
\label{eq:Hamiltonian}
H=-\sum_{\langle
ij,i^{\prime}j^{\prime}\rangle}t_{ij,i^{\prime}j^{\prime}}C_{ij}^{\dag}
C_{i^{\prime}j^{\prime}}+\mu\sum_{ij}C_{ij}^{\dag} C_{ij},
\end{equation}
whereas $C_{ij}^{\dag}$ ($C_{ij})$ is the operator of creating (annihilating) one electron at a lattice site with site coordinates being $x_i$ and $y_j$ respectively, $\langle , \rangle$ denotes the nearest-neighbors, $t_{ij,i^{\prime}j^{\prime}}$ is the nearest-neighbor hopping integral, and $\mu$ is the chemical potential, i.e. the Fermi level, which can be adjusted by an effective gate voltage directly applied on the graphene ribbon.

Figure \ref{fig:structure} schematically shows an armchair-shaped graphene ribbon (AGR), connected with two semi-infinite square-lattice quantum wires described also by a tight binding Hamiltonian with only the nearest-neighbor hopping. There are $N$(length)$\times$$M$(width) lattice sites in the AGR. For simplicity, the nearest-neighbor hopping integral $t_{ij,i^{\prime}j^{\prime}}$ in the AGR sets to $t_0$, adopted as an energy unit in this article. We further assume the hopping integrals in both the left and right leads and the interface hopping integrals between the leads and AGR all being $t_0$. We use the natural open boundary condition in the calculations, which means that there are no longer dangling bonds along the boundaries, equivalent to the saturation by hydrogens in experiments.


For the Shr\"{o}dinger equation $\hat H\psi(E)=E\psi(E)$ of the considered system (Fig. \ref{fig:structure}), any wavefunction $\psi(E)$ at a given energy $E$ can be expressed by a linear combination of localized Wannier bases $|ij\rangle =C_{ij}^{\dag}|vaccum\rangle$, that is, $\psi(E)=\sum_{ij}\alpha_{ij}|ij\rangle$ with the complex coefficients $\alpha_{ij}$ to be determined. In other words, a set of $\{\alpha_{ij}\}$ is the site representation of the wave function. In the left or right lead, $\alpha_{ij}$ is further denoted as $\alpha^{L}_{ij}$ or $\alpha^R_{ij}$.

In the AGR, by applying the Hamiltonian (\ref{eq:Hamiltonian}) on $\psi(E)=\sum_{ij}\alpha_{ij}|ij\rangle$ we obtain the following equation regarding the wavefunction in the scattering region for a given energy $E$,
\begin{equation}\label{eq:coefficient}
E\alpha_{ij}=-t_0\sum_{\rho,\delta}\alpha_{i+\rho,j+\delta}+\mu \alpha_{ij},
\end{equation}
where $\rho$ and $\delta$ denote the nearest neighbors along $x$ and $y$ directions respectively.


We now define a column vector $\hat{\alpha_i}$ which consists of all the $M$ $\alpha$-coefficients with the same $x$-axis index $i$,
\begin{equation}\label{eq:vector}
\hat{\alpha}_i=
\left(
\begin{array}{ccc}
 \alpha_{i1}  \\
 \alpha_{i2} \\
 \vdots \\
 \alpha_{iM}
\end{array}
\right).
\end{equation}
After rearranging, we can then rewrite Eq.~\eqref{eq:coefficient} in a more compact form as
\begin{equation}\label{eq:transfer matrix}
\left(
\begin{array}{ccc}
 \hat{\alpha}_{i-1}  \\
 \hat{\alpha}_i
\end{array}
\right)
=\hat{\chi}_i
\left(
\begin{array}{ccc}
 \hat{\alpha}_i  \\
 \hat{\alpha}_{i+1}
\end{array}
\right).
\end{equation}
Here $\hat{\chi}_i$ is the so-called $i$-th transfer matrix which elements consist of $E$, $t_0$, and $\mu$. As its name means, $\hat{\chi}_i$ connects $M$ $\alpha$-coefficients of any slice $i$ with $M$ $\alpha$-coefficients of its two neighbor slices $i-1$ and $i+1$. There are totally $N$ transfer matrices in the AGR (Fig. 1).

In each lead, there are $M$ right-traveling waves (channels) and $M$ left-traveling waves (channels) for a given energy $E$, respectively. Each channel is defined by the corresponding
transverse wave vector $k_y^n$ determined by the open boundary condition, namely forming standing waves, $k_{y}^{n}=\frac{n \pi}{M+1}$ with $n$ being an integer from
$1$ to $M$ and the lattice constant $a$ being assumed as a length unity. Physically when an unity-amplitude right-traveling wave in the $n$-th channel is scattered into the $n^{\prime}$-th channel, the wavefunction in the left and right semi-infinite leads can be expressed respectively \cite{transfer} as
\begin{equation}\label{wave}
\left\{
  \begin{array}{ll}
    \alpha^{L}_{n,ij}= & \sum\limits_{n^{\prime}}
    \left(\delta_{n^{\prime}n}e^{i k_{x}x_i}+r_{n^{\prime},n}e^{-i
    k_{x}^{\prime}x_i}\right)\sin(k_{y}^{n^{\prime}}y_j), \\
    \alpha^{R}_{n,ij}= & \sum\limits_{n^{\prime}} t_{n^{\prime},n}e^{i
    k_{x}^{\prime}x_i}\sin(k_{y}^{n^{\prime}}y_j),
  \end{array}
\right.
\end{equation}
where $r_{n^{\prime},n}$ and $t_{n^{\prime},n}$ are the reflection and transmission coefficients from the $n$-th to the $n^{\prime}$-th channel respectively, and the continuous longitudinal wave vector $k_{x}$ and the discrete transverse wave vector $k_{y}^{n}$ of the $n$-th channel satisfy the following dispersion relation of the square lattice tight-binding model,
\begin{equation}\label{eq:lead dispersion}
E= -2t_0(\cos k_{x}+\cos k_{y}^{n}),
\end{equation}
which uniquely determines $k_x$ in the $n$-th channel for a given energy $E$, thus denoted as $k_x^n$ from now.

As shown in Fig. 1, at the interfaces between the AGR and the leads the lead wavefunction represented by Eq. \eqref{wave} naturally extend to the site columns indexed with 0 and 1 from the left side and the site columns indexed with $N+1$ and $N$ from the right side respectively. Then by using the column vector notation (Eq. \eqref{eq:vector}), we can rewrite the $n$-th channel wavefunction at the interface more compactly as,
\begin{equation}\label{eq:r t matrix}
\begin{array}{ccc}
\left(
\begin{array}{ccc}
 \hat{\alpha}_0  \\
 \hat{\alpha}_1
\end{array}
\right)
=\hat{R}\hat{r}_{n}+\hat{\delta}_{n} & \text{and} &
\left(
\begin{array}{ccc}
 \hat{\alpha}_N  \\
 \hat{\alpha}_{N+1}
\end{array}
\right)
=\hat{T}\hat{t}_{n},
\end{array}
\end{equation}
where
\begin{equation}
\label{eq:r t vector}
\begin{array}{ccc}
\hat{r}_{n}=
\left(
\begin{array}{ccc}
r_{1,n} \\
r_{2,n} \\
\vdots \\
r_{(M-1),n} \\
r_{M,n}
\end{array}
\right) & , &
\hat{t}_{n}=
\left(
\begin{array}{ccc}
t_{1,n} \\
t_{2,n} \\
\vdots \\
t_{(M-1),n} \\
t_{M,n} \\
\end{array}
\right)
\end{array}
\end{equation}
and
\begin{equation}
\label{eq:delta vector}
\hat{\delta}_{n}=
\left(
\begin{array}{ccc}
e^{ik_{x}^{n}x_{0}}\sin(k_{y}^{n}y_{1}) \\
\vdots \\
e^{ik_{x}^{n}x_{0}}\sin(k_{y}^{n}y_{M}) \\
e^{ik_{x}^{n}x_{1}}\sin(k_{y}^{n}y_{1}) \\
\vdots \\
e^{ik_{x}^{n}x_{1}}\sin(k_{y}^{n}y_{M})
\end{array}
\right).
\end{equation}
For convenience to represent matrices $\hat{R}$ and $\hat{T}$ with dimensions $2M \times M$, we further define two $M\times M$ matrices as follows,
\begin{equation}
\label{eq:x function}
\hat{\xi}(x_{i})=
\left(
\begin{array}{ccc}
e^{ik_{x}^{1}x_{i}} & & \\
& \ddots & \\
& & e^{ik_{x}^{M}x_{i}}
\end{array}
\right)
\end{equation}
and
\begin{equation}
\label{eq:y fuction}
\hat{\zeta}=
\left(
\begin{array} {ccc}
\sin(k_{y}^{1}y_{1}) &
\cdots &
\sin(k_{y}^{M}y_{1}) \\
\vdots & \vdots & \vdots \\
\sin(k_{y}^{1}y_{M}) &
\cdots &
\sin(k_{y}^{M}y_{M}) \\
\end{array}
\right).
\end{equation}
With the above definition, matrices $\hat{R}$ and $\hat{T}$ can be represented by the products of $\hat{\xi}$ and $\hat{\zeta}$ as follows,
\begin{equation}
\label{eq:R matrix}
\hat{R}=
\left(
\begin{array}{ccc}
\hat{\zeta} & 0 \\
0 & \hat{\zeta}
\end{array}
\right)
\left(
\begin{array}{ccc}
\hat{\xi}(x_0)^{*}\\
\hat{\xi}(x_1)^{*}
\end{array}
\right)
\end{equation}
and
\begin{equation}
\label{eq: T matrix}
\hat{T}=
\left(
\begin{array}{ccc}
\hat{\zeta} & 0 \\
0 & \hat{\zeta}
\end{array}
\right)
\left(
\begin{array}{ccc}
\hat{\xi}(x_N)  \\
\hat{\xi}(x_{(N+1)})
\end{array}
\right),
\end{equation}
where $*$ means the complex conjugation.

According to Eq. \eqref{eq:transfer matrix} conventionally one multiplies all $N$ transfer matrices $\hat{\chi}_i$$(i=1,\ldots,N)$ to establish a direct connection between the left interface and right interface across the scattering region for the wavefunction,
\begin{equation}
\left(
\begin{array}{ccc}
 \hat{\alpha}_{0}  \\
 \hat{\alpha}_1
\end{array}
\right)
=\hat{\chi}_{1}\hat{\chi}_{2}\cdots\hat{\chi}_{N-1}\hat{\chi}_{N}
\left(
\begin{array}{ccc}
 \hat{\alpha}_N  \\
 \hat{\alpha}_{N+1}
\end{array}
\right),
\end{equation}
which determines the reflection and transmission coefficients in combination with Eqs. \eqref{eq:r t matrix}. However, the $N$-fold multiplication of the transfer matrix will bring out the aforementioned numerical instability for a large $N$, which has been found for $N\geq 10$ with $M\geq 50$ in our calculations.

\begin{figure}[tbh]
\begin{center}
\includegraphics[width=8.5cm]{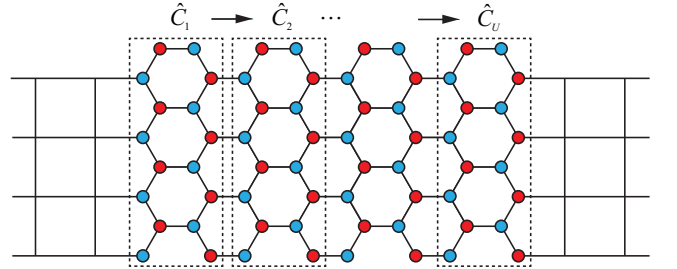}
\caption{(Color online) Transfer matrix renormalization scheme. The whole system is divided into $U$ subunits which contains $S$ columns. Here $U=N/S$. Notice that the last subunit may have less columns than $S$, when $N$ is not integral multiple of $S$.}
\label{fig:renormalization scheme}
\end{center}
\end{figure}

In order to solve the numerical instability, we propose the transfer matrix renormalization scheme in which we first divide the scattering region into $U$ subregions, each of which contains $S$ columns, as shown in Fig.~\ref{fig:renormalization scheme}. For all the subregions, we successively take $S$-fold multiplication of the transfer matrix according to Eq. \eqref{eq:transfer matrix} to establish direct connections between the interfaces across the subregions for the wavefunction, represented by
\begin{equation}
\label{eq:common re}
\left(
\begin{array}{ccc}
 \hat{\alpha}_{(i-1)S}  \\
 \hat{\alpha}_{(i-1)S+1}
\end{array}
\right)
=\hat{C}_{i}
\left(
\begin{array}{ccc}
 \hat{\alpha}_{iS}  \\
 \hat{\alpha}_{iS+1}
\end{array}
\right), \\
\end{equation}
where $\hat{C}_i=\hat{\chi}_{(i-1)S+1}\hat{\chi}_{(i-1)S+2}\cdots
\hat{\chi}_{iS}, ~i=1,\cdots,~U$. Then in combination with Eqs. \eqref{eq:r t matrix}, we lump all the equations represented by Eq. \eqref{eq:common re} into a following set of linear equations containing the reflection and transmission coefficients from a fixed (the $n$-th) channel respectively to all the $M$ channels plus the wavefunction values at the interfaces between the subregions as the unknowns,
\begin{equation}
\label{eq:renormalization eq}
\hat{A}\hat{\lambda}_{n}=\hat{B}_{n},
\end{equation}
with
\begin{equation}
\label{eq:A matrix}
\hat{A}=
\left(
\begin{array}{ccccc}
-\hat{R} & \hat{C}_1 & & & \\
& -\hat{I} & \hat{C}_2 & & \\
& & \ddots & \ddots & \\
& & & -\hat{I} & \hat{C}_U\hat{T} \\
\end{array}
\right),
\end{equation}
\begin{equation}
\label{eq:lambda and B}
\begin{array}{ccc}
\hat{\lambda}_{n}=
\left(
\begin{array}{ccc}
\hat{r}_{n} \\
\hat{\alpha}_S  \\
\hat{\alpha}_{S+1} \\
\hat{\alpha}_{2S}\\
\hat{\alpha}_{2S+1}\\
\vdots \\
\hat{\alpha}_{(U-1)S}\\
\hat{\alpha}_{(U-1)S+1}\\
\hat{t}_{n} \\
\end{array}
\right)
& \text{and} &
\hat{B}_{n}=
\left(
\begin{array}{c}
\hat{\delta}_{n} \\
0
\end{array}
\right).
\end{array}
\end{equation}
In Eq.~\eqref{eq:A matrix}, the other elements not listed in matrix $\hat{A}$ are all zero. Clearly we will first make such a division in practice that the numerical instability will not take place in matrix $\hat{C}_i$, which can be easily realized once the subregions are short enough. On the other hand, we would also like to make the subregions as long as possible to reduce the size of matrix $\hat{A}$ as small as possible. So it needs to take a balance in making division.

Eq. \eqref{eq:renormalization eq} now becomes the core of the whole problem. As long as we solve Eq. \eqref{eq:renormalization eq}, we can obtain the reflection and transmission coefficients of the $n$-th channel to all the $M$ channels respectively. Then we use the Landauer formula to calculate conductance,
\begin{equation}
\label{eq:landauer formula Fano Factor}
\begin{array}{ccc}
G=\frac{2e^2}{h}tr(\tilde{t}\tilde{t}^{\dag}) &  &
\end{array}
\end{equation}
with
\begin{equation}
\label{eq:tt}
\tilde{t}=
\left(
\begin{array}{ccccc}
\eta_{1,1}t_{1,1} & \cdots & \eta_{1,M}t_{1,M} \\
\vdots & \vdots & \vdots \\
\eta_{M,1}t_{M,1} & \cdots & \eta_{M,M}t_{M,M} \\
\end{array}
\right)
\end{equation}
and
\begin{equation}
\label{eq:eta}
\eta_{n^{\prime},n}=
\left\{
\begin{array}{cccccc}
\sqrt{\frac{|\sin k_x^{n^{\prime}}|}{|\sin k_x^{n}|}} & , &
\text{for real nonzero} & k_x^{n} & \text{and} & k_x^{n^{\prime}}; \\
0 & , & \text{otherwise}. & & & \\
\end{array}
\right.
\end{equation}

The renormalized transfer matrix method proposed here can be easily generalized to the case of an irregular graphene nanoribbon composed by a series of nanoribbons with different width \cite{Zheng-GTM}. Similar to Eqs. \eqref{eq:renormalization eq} and \eqref{eq:A matrix}, a set of the linear equations can be constructed as well, in which however the dimensions of the matrices $\hat{C}_i$ are variant and depend on the width of the local graphene nanoribbon.

Finally we comment on the application of renormalized transfer matrix method on some relevant structures. Our method is not applicable to deal with graphene nanoribbon lead only because the wavefunction in graphene nanoribbon lead cannot be expressed analytically as that in quantum wire lead. For other relevant structures, the coefficients of wavefunctions for two columns of lead lattice sites adjacent to lead/graphene interfaces are usually included, and they are related with transmission matrix $T_{i}$ and reflection matrix $R_{i}$ in lead $i$. When the transverse size of graphene devices is different from that of leads and/or there are multi-terminals, transfer matrix method is not applicable and the linear equations for those coefficients of wavefunctions for all lattice sites adjacent to lead/graphene interfaces are obtained directly from Sch$\ddot{o}$rdinger equations. By applying renormalized transfer matrix method for the other uniform sub-systems combined with above linear equations, we obtain the matrix A analogy to that in Eq. \ref{eq:A matrix} and sparse too. It is easy to apply RTMM to investigate electronic transport through zigzag graphene between quantum wire leads, and the only difference lies in the specific form of transfer matrix. For $\chi_{i}$ in armchair graphene nanoribbon had been given \cite{Hu}, while $\chi_{i}$ in zigzag graphene nanoribbon is listed in the note \cite{note}.

\section{Optimized Guass elimination scheme}

The efficiency of calculating the whole problem depends on how to solve the set of linear equations represented by Eqs. \eqref{eq:renormalization eq}, \eqref{eq:A matrix}, and \eqref{eq:lambda and B}, in which $\hat{A}$ is a $2UM\times2UM$ block matrix. In general, there are two standard algorithms to solve a set of linear equations, i.e. Gaussian elimination and $LU$ decomposition\cite{recipes}. Both of them consist of $\frac{1}{3}(2UM)^3$ loop executions (each loop containing one subtraction and one multiplication), where $2UM$ is the dimension of the coefficient matrix in a set of linear equations, and can be operated in place to save memory. If the coefficient matrix is a sparse matrix, these two methods can be modified based on the characteristics of its sparseness to greatly improve the performance.

In the case of matrix $\hat{A}$, clearly the Gaussian elimination method exploits the sparse structure more easily than the $LU$ decomposition. However the conventional full pivoting, designed in the Gaussian elimination method to reduce computing roundoff errors, has to be given up since it picks up a pivoting element among all the matrix elements so as to mess up the structure of matrix $\hat{A}$. On the other hand, it is well-known that the Gaussian elimination method without proper pivoting is unreliable. For matrix $\hat{A}$ we notice that the nonzero elements are uniformly distributed except in sub-matrices $\hat{R}$ and $\hat{T}$ which describe the two interfaces. This indicates that the full pivoting is unnecessary for matrix $\hat{A}$.





\begin{figure}[tbh]
\begin{center}
\includegraphics[width=8.0cm]{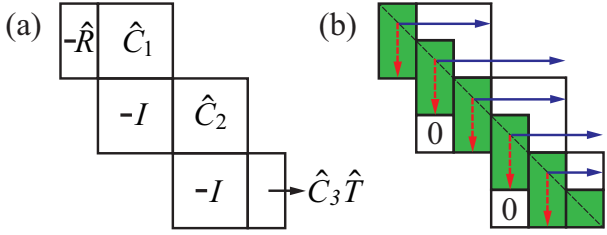}
\caption{(Color online) (a) Block matrix $\hat{A}$ for $U=3$, only non-zero
elements are listed. (b) Schematic local maximum pivoting Gaussian elimination method (LMPG),
where green rectangle and the last square represent doing maximum pivoting in these local areas,
blue solid and red dash rows show the range of normalization and elimination, respectively.
Black dash line donates the diagonal line.}
\label{fig:local}
\end{center}
\end{figure}





Targeting matrix $\hat{A}$, we develop a local maximum pivoting Gaussian elimination method (LMPG), in which a pivoting can be well undertaken in a local area to realize the same reduction of computing roundoff errors as the full pivoting. Figure 3 schematically shows the areas for such a local pivoting and the ranges for subsequent normalization and elimination respectively. The corresponding algorithm is formulated in Alg.~\ref{alg:gauss}, in which two important functions $f(k,p)$ and $g(k,q)$ are introduced to specify the areas and ranges for pivoting, elimination and normalization respectively. We now describe these two functions. Firstly matrix $\hat{A}$ represented by Eq. \eqref{eq:A matrix} can be divided into $2U\times 2U$ sub-blocks, each of which contains $M\times M$ elements. We then denote the positions of these sub-blocks in matrix $\hat{A}$ by a pair of integers $(p,~q)$, where $p,~q=1,\cdots,~2U$. If $k$ stands for the row index of matrix $\hat{A}$, each $k$ will correspond to a pair of $(p,~q)$. Thus the two functions $f(k,p)$ and $g(k,q)$ are defined as follows,
\begin{equation}
\label{eq:f}
f(k,p)=
\left\{
\begin{array}{lll}
2UM & , & p=2U, \\
(p+1)M & , & \text{otherwise,}
\end{array}
\right.
\end{equation}
and
\begin{equation}
\label{eq:g}
g(k,q)=
\left\{
\begin{array}{lll}
[\frac{q}{2}] \times2M+3M & , & q\leq2U-3, \\
2UM & , & \text{otherwise},
\end{array}
\right.
\end{equation}
where $[\frac{q}{2}]$ means taking the integer part of $\frac{q}{2}$.

By examining Eqs. \eqref{eq:renormalization eq}, \eqref{eq:A matrix}, and \eqref{eq:lambda and B}, we further notice that even though the constant column vectors $\hat{B}_n$ in the right side of the equations are different with each other, the coefficient matrices $\hat{A}$ are identical for all the $M$ incident channels. Therefore all the $M$ channels can be simultaneously dealt with at one time rather than successive $M$ times, by reformulating Eqs. \eqref{eq:renormalization eq}, \eqref{eq:A matrix}, and \eqref{eq:lambda and B} into the following composite set of equations,
\begin{equation}
\hat{A}(\hat{\lambda}_1,\cdots,~\hat{\lambda}_M)=(\hat{B}_1,\cdots,~\hat{B}_M).
\end{equation}


The comparison on efficiency between the local maximum pivoting Gaussian elimination method and the standard Gaussian elimination method is summarized in Table~\ref{table:efficiency}. Here the computational complexity and memory requirement are quantitatively analyzed by the times of loop executions (TL) (each loop containing one subtraction and one multiplication) and the numbers of matrix elements to store (NE), respectively. It turns out that TL and NE are greatly reduced from O($U^3M^{3}$) and O($U^{2}M^{2}$) to O($UM^{3}$) and O($M^{2}$), respectively. For example, in the case of $S=10$, TL and NE by LMPG for a system with lattice sites as large as $1000 \times 1000$ is 6-order and 3-order less than those by CCPG, respectively.

\begin{algorithm}[h]
\caption{Local maximum Gaussian elimination method}
\label{alg:gauss}
\begin{algorithmic}[1]
\FOR{each $k\in [1,2UM-1]$}
\STATE \quad \textbf{Local pivoting} \\
\STATE \qquad $\max|a_{ij}|, ~k\leq i\leq f(k,p),~k\leq j\leq qM$ \\
\STATE \qquad suppose $\max|a_{ij}|=a_{uv}$ \\
\STATE \quad \textbf{Two rows exchange}  \\
\STATE \qquad $a_{kj}\Leftrightarrow a_{uj}, ~j=k,\cdots,~g(k,q)$ \\
\STATE \quad \textbf{Two columns exchange}  \\
\STATE \qquad $a_{ik}\Leftrightarrow a_{iv}, ~i=k,\cdots,~f(k,p)$ \\
\STATE \quad \textbf{Normalization} \\
\STATE \qquad $a_{kj}/a_{kk}\Rightarrow a_{kj}, ~j=k+1,\cdots,~g(k,q)$ \\
\STATE \qquad $b_{kj}/a_{kk}\Rightarrow b_{kj}, ~j=1,\cdots,~M$ \\
\STATE \quad \textbf{Elimination} \\
\STATE \qquad $a_{ij}-a_{ik}a_{kj}\Rightarrow a_{ij}$ \\
\STATE \qquad $\qquad i=k+1,\cdots,~f(k,p), ~j=k+1,\cdots,~g(k,q)$ \\
\STATE \qquad $b_{ij}-a_{ik}b_{kj}\Rightarrow b_{ij}$ \\
\STATE \qquad $\qquad i=k+1,\cdots,~f(k,p), ~j=1,\cdots,~M$ \\
\ENDFOR
\STATE \textbf{Solution} \\
\STATE \quad $b_{(2UM)j}/a_{(2UM)(2UM)}\Rightarrow \lambda_{(2UM)j}, ~j=1,\cdots,~M$ \\
\FOR{each $k\in [2UM-1,1]$}
\STATE \quad \textbf{Back substitution} \\
\STATE \qquad $b_{kj}-\sum_{i=k+1}^{g(k,q)}a_{ki}\lambda_{ij}\Rightarrow \lambda_{kj}, ~j=1,\cdots,~M$ \\
\IF{mod(k-1,M)=0}
\STATE \quad \textbf{Recover} $\lambda_{ij}, ~i=(p-1)M+1,\cdots,pM,~j=1,\cdots,M$ \\
\ENDIF
\ENDFOR
\end{algorithmic}
\end{algorithm}

\begin{table}
\caption{A efficiency contrast between conventional column pivoting Gaussian
elimination (CCPG) and local maximum pivoting
Gaussian elimination method (LMPG). Computational complexity and memory requirement
are analyzed by the times of loop executions (TL) (each loop containing one
subtraction and one multiplication) and the numbers of matrix elements to store (NE),
respectively. Assuming
$S=10$, we show the difference of two algorithms for different size systems.
If the matrix elements are declared as double precision complex numbers, the
memory used in solving the equations are listed in the table.}
\label{table:efficiency}
\begin{tabular}{|c|c|c|c|c|}
  \hline
  System & \multicolumn{2}{c|}{TL}
  &\multicolumn{2}{c|}{NE} \\
  \cline{2-5}
  Size & CCPG & LMPG & CCPG & LMPG \\
  \cline{1-5}
  M N & $2.67U^3M^3$ & $12UM^3$ & $4U^2M^2$ & $32M^2$ \\
  \cline{1-5}
  100 100 & $2.67\times10^{9}$ & $1.2\times10^8$ & $61$~MB & $4.88$~MB \\
  \cline{1-5}
  1000 1000 & $2.67\times10^{18}$ & $1.2\times10^{12}$ & $596$~GB & $0.477$~GB \\
  \hline
\end{tabular}
\end{table}

As we see, by utilizing the sparse structure of matrix $\hat{A}$, we can drastically reduce the computational memory as well as the numbers of loop executions, which is vital for us to being capable of calculating a large system.


\section{armchair graphene nanoribbons}

\begin{figure}[tbh]
\begin{center}
\includegraphics[width=8.6cm]{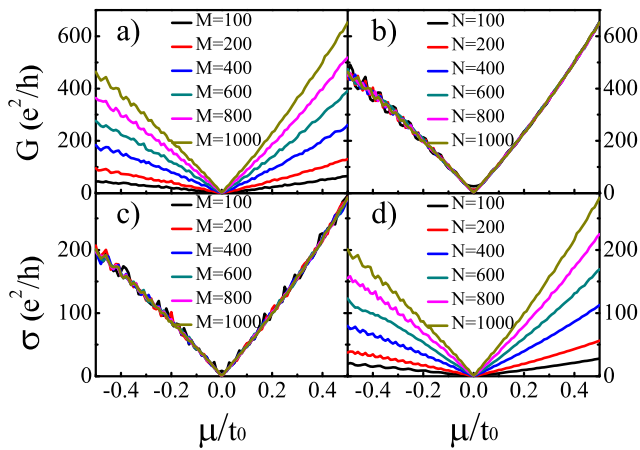}
\caption{(Color online) Transport properties of pure AGR. (a) and (b) are
conductance versus chemical potential for fixed $N$ ($N=1000$) and
fixed $M$ ($M=1000$), respectively. (c) and (d) are conductivity versus
chemical potential for fixed $N$ ($N=1000$) and fixed $M$ ($M=1000$),
respectively.}
\label{fig:AGR}
\end{center}
\end{figure}

Previously, we applied conventional transfer matrix to investigate the transport in
small-scale graphene-based system not exceeding $O(10^{4})$ lattice sites \cite{ZhangGP-01,ZhangGP-02,ZhangGP-03}.
For small graphene-based system, it is difficult to extract basic transport properties as shown in Fig. 3 \cite{ZhangGP-03}, therefore transport in large-scale graphene-based system is highly desired.
Meanwhile, the transport in a large-scale graphene up to $O(10^{6})$ lattice sites was investigated by diagonalizing transfer matrix \cite{Hu}, which is powerful to deal with uniform and pure system while difficult to solve disordered system and/or system with impurities.
Here we study the transport properties of AGR connected to normal
leads, with a variety of widths and lengths at different chemical potential
(i.e. different gate voltages) and the results are summarized in
Fig.~\ref{fig:AGR}.
Physically, there is a small
energy gap for a finite and semiconducting graphene ribbon, while at a finite $\mu$ over the
energy gap, the ballistic transport is thus expected in pure system since there are always
a number of channels for electrons to propagate through. Therefore, for a fixed width with different lengths the curves of the conductance $G$ versus $\mu$ coincides exactly except the oscillation due to quantum interference
as shown in Fig.~\ref{fig:AGR}(b), in which $G$ is independent of $L$. On the other hand, for a fixed length with different widths, the conductance $G$ is proportional to the width $W$ as shown in Fig.~\ref{fig:AGR}(a).
It turns out that the conductivity of AGR
$\sigma$ being $GL/W$ merge together for a fixed length with different
widths as shown in Fig.~\ref{fig:AGR}(c) and $\sigma$ is proportional
linearly to the length $L$, for ribbons with the same $W$  as shown in Fig.~\ref{fig:AGR}(d),
respectively. Furthermore, $\sigma$ is superlinearly
to the chemical potential $\mu$, but with the different slopes between the positive
and negative chemical potential in AGR, as shown in Fig.~\ref{fig:AGR}. This also
means that the AGR conductivity increases with the carrier density, and the
asymmetrical behavior between electrons and holes, due to the occurrence of odd-numbered rings
formed at the interface of AGR and normal metal contacts \cite{ZhangGP-03}, is consistent with the experimental observation \cite{exp0,exp1}. Finally we compare the transport in small and large graphene-based materials and find that the transport in large system is close to experimental observations, since the sizes of samples in experiments usually reach microns. However, the breaking of electron-hole symmetry in transport exists in both small and large armchair-shaped graphene system.

\begin{figure}[tbh]
\begin{center}
\includegraphics[width=8.6cm]{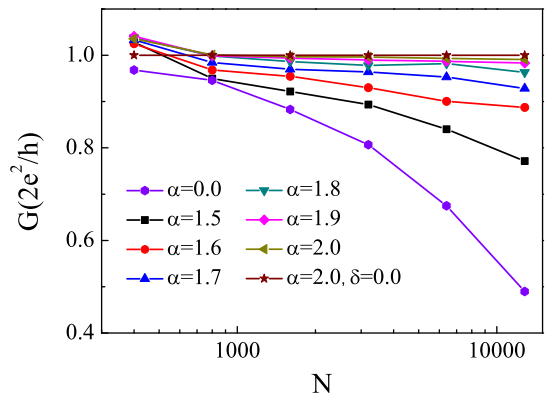}
\caption{(Color online) Transport properties of long range correlated disordered AGR as $\alpha$ varies from 0 to 2.0, where $M=52$, $\delta=0.1$ and $\mu=0$. $\alpha=0$ and $\delta=0$ corresponds to Anderson disorder and no disorder respectively.}
\label{fig:LRCD}
\end{center}
\end{figure}

In graphene samples, disorder and impurities are usually inevitable. Here we study the effect of Anderson disorder with long-range correlation on the transport in large graphene at charge neutral point.
It was commonly believed that Anderson localization \cite{Anderson-PR1958} in one- and two-dimensional systems is induced by even a weak disorder from the scaling theory \cite{Anderson-PRL1979}. However metal-insulator transition occurs in low-dimensional disordered system with long-range correlation \cite{FABF-longPRL1998,Izrailev-RPL1999,Zhang-1,Cheraghchi-PRB2011,disorder-correlation}. The metallic phase occurs since the relative disorder between any two lattice sites decreases for strongly correlated disorder \cite{Zhang-1}. The long-range correlated disordered onsite energies are shown in Ref. \onlinecite{disorder-correlation}, which change from random to smooth and striped when the correlation parameter $\alpha$ increases from 1.0 to 2.5. $\alpha=0$ corresponds to uncorrelated disorder, i.e., Anderson disorder. For narrow AGRs (e.g., $M=10$) with increasing lengths, the conductance either approaches to $2e^{2}/h$ as $\alpha \ge 1.86$ or $G \propto \exp(-aN)$ as $\alpha=0.1$ (at $\delta=0.1$) \cite{disorder-correlation}, implying that metal-insulator transition is induced by long-range correlated disorder.

Since the localization length in graphene increases with the width as a result of more channels in wider graphene, the length of AGR should increase till comparable to the localization length in order to study metal-insulator transition. Therefore the application of RTMM to large-scale graphene system is necessary in this case.
It is expected that for fixed strength of disorder, metal-insulator transition induced by the strength of correlation, denoting by two different scaling behaviors as above, takes place when the length of AGRs increases.
In Fig.~\ref{fig:LRCD}, the conductance depends on the length $N$ (varying from 400 to 12800) when $\alpha$ changes from 1.5 to 2.0, $\delta=0.1$ and $M=52$. For clear comparison, the conductances of AGR without any amount of disorder and with Anderson disorder are also shown as two extremes. It is found that the conductance in pure AGR equals to $2e^{2}/h$ as $N$ is larger than 400. On the other hand, the conductance in AGR with Anderson disorder decreases monotonically as $N$ increases and finally decays exponentially in even longer AGR. In the presence of long-range correlated disorder, the conductances in AGRs (averaged between 500 samples) are between above two extremes and the conductance increases with the strength of correlation. As $N$=400 and $\alpha\ge 1.5$, the average conductance is a little higher than that in pure graphene due to the deviation of the conductance. The conductance curves eventually collapse and approach that in pure AGR as $\alpha$ is larger than 1.8, and the conductance decreases as the length increases otherwise. Therefore the long-range correlation of disorder induces the localization-delocalization transition in AGR.

\section{Conclusion}

Renormalized transfer matrix method (RTMM) is proposed and the computational speed and memory usage have been greatly improved by optimization. RTMM is used to study the electronic transport in large scale pure and long-range correlated disordered armchair graphene nanoribbon (with carbon atoms up to $10^{6}$ for pure case) between quantum wire contacts. As for pure AGR, the conductance is superlinear with the Fermi energy and the conductance of ballistic transport
is linear with the width while independent of the length. As for disordered AGR with long-range correlation,
there is metal-insulator transition induced by the correlation strength of disorder.
It is straightforward to extend RTMM to investigate transport in large scale system with irregular structure.


\begin{acknowledgements}

This work is supported by NSF of China (Grant Nos. 11004243, 11190024, and 51271197), National Program for Basic Research of MOST of China (Grant No. 2011CBA00112). Computational resources have been provided by the Physical Laboratory of High Performance Computing in RUC.

\end{acknowledgements}


\begin{references}
\bibitem{note} $\chi_{i}$ (with dimension $2M \times 2M$ and $M$ being the width of ZGR) is divided into four 
$M \times M$sub-blocks, i.e., $\chi_{i}^{(1,1)}$, $\chi_{i}^{(1,2)}$, $\chi_{i}^{(2,1)}$ and $\chi_{i}^{(2,2)}$. The nonzero elements of $\chi_{i}^{(1,1)}$ are $\chi_{i}^{(1,1)}(k,k)=E-\mu_{i,k}$, $\chi_{i}^{(1,1)}(k,k+1)=-1.0$ when $i+k$ is odd, and $\chi_{i}^{(1,1)}(k,k-1)=-1.0$ when $i+k$ is even. $\chi_{i}^{(1,2)}=-I$, $\chi_{i}^{(2,1)}=I$ and $\chi_{i}^{(2,2)}=0$.
\end{references}
\end{document}